\documentclass[aps,twocolumn,showpacs]{revtex4}
\errorstopmode

\usepackage{graphicx}
\usepackage{amsmath}
\usepackage{amssymb}
\newcommand\eps\varepsilon
\newcommand{\dd}{{\rm d}}

\newcommand{\fii}{\phi}
\newcommand{\e}{{\rm e}}

\begin{document}

\title{Approximate magnifying perfect lens with positive refraction}

\author{Tom\'a\v{s} Tyc}\author{Martin \v Sarbort}
\affiliation{Institute of Theoretical Physics and Astrophysics,
Masaryk University, Kotl\'a\v rsk\'a 2, 61137 
Brno, Czech Republic}

\date{\today}
\begin{abstract}
We propose a device with a positive isotropic refractive index that creates an
approximate magnified perfect real image of an optically homogeneous
three-dimensional region of space within geometrical optics. Its key ingredient
is a new refractive index profile that can work as an approximate perfect lens
on its own, having a very moderate index range.  \\ 
\end{abstract}
\maketitle

\newpage

In perfect imaging, light rays emerging from any point P of some
three-dimensional region are perfectly (stigmatically) reassembled at another
point P', the image of P.  Perfect imaging has been one of the hot topics in
modern optics since 2000 when J. Pendry showed that a slab of a material with
negative refractive index~\cite{Pendry2000} can work as a perfect imaging
device (perfect lens).  Much effort has then been put into designing and
constructing perfect lenses based on materials with negative refractive
index~\cite{Shalaev2007-review}, which led e.g.~to demonstrations of
sub-wavelength resolution~\cite{Fang2005-neg_n-silver_imaging}. 

On the other hand, as early as in 1854 J. C. Maxwell found a device with an
isotropic and positive refractive index that images the whole space perfectly,
which he called fish eye.  More than 150 years later, U. Leonhardt and
T. Philbin showed that Maxwell's fish eye provides perfect imaging not only in
terms of geometrical optics also in the full framework of wave
optics~\cite{Ulf2009-fisheye,Ulf2010-fisheye}, and therefore enables
sub-wavelength resolution similarly as perfect lenses based on negative
refraction. Recent experiments have confirmed this~\cite{Ma2010-fisheye}.
Only a few other perfect lenses with an isotropic positive refractive index
were known until recently.  Even less was known about devices that would image
perfectly homogeneous regions of space, i.e., regions with a uniform refractive
index. Indeed, even in the last issue of Born and Wolf's Principles of
Optics~\cite{BornWolf} we read that the only known example of such a device is
a plane mirror or a combination thereof.  This has been changed by a recent
excellent work of J. C. Mi\~nano~\cite{Minano2006} who proposed several new
perfect lenses imaging homogeneous regions and also showed that some well-known
optical devices such as Eaton lens or Luneburg lens~\cite{Luneburg1964} are in
fact perfect lenses as well. All of them have unit magnification, giving an
image of the same size as the original object.

Here we present a lens that provides an approximate perfect real image of a
homogeneous region of 3D space with an arbitrary magnification. Our device is a
non-trivial combination of Maxwell's fish eye and a new refractive index
profile. This profile equipped with a spherical mirror can work as an
approximate perfect lens on its own, giving a real image of a homogeneous
sphere and using just a moderate refractive index range. Our lens employs
isotropic material with positive refractive index.

\begin{figure}
\begin{center}
\includegraphics[width=8cm]{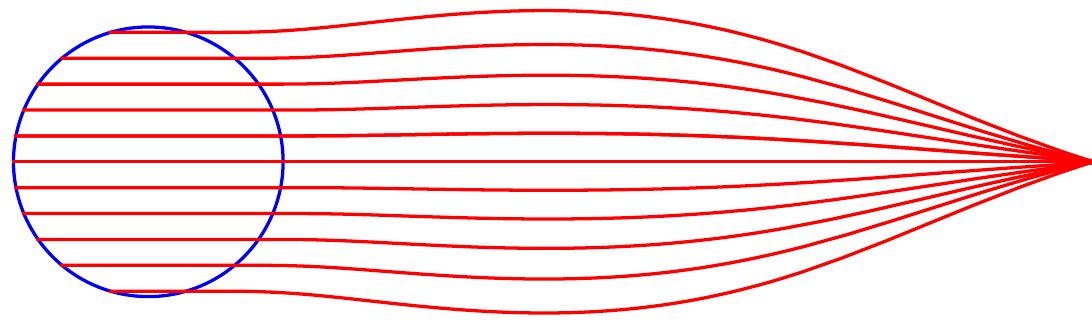}
\end{center}
\caption{Medium with a spherically-symmetric refractive index that focuses
  parallel rays inside an optically homogeneous unit sphere (blue) to a point
  at distance $R$ from its center (here $R=7$).}
\label{rays}
\end{figure}

\begin{figure}
\begin{center}
\includegraphics[angle=90,width=6.5cm]{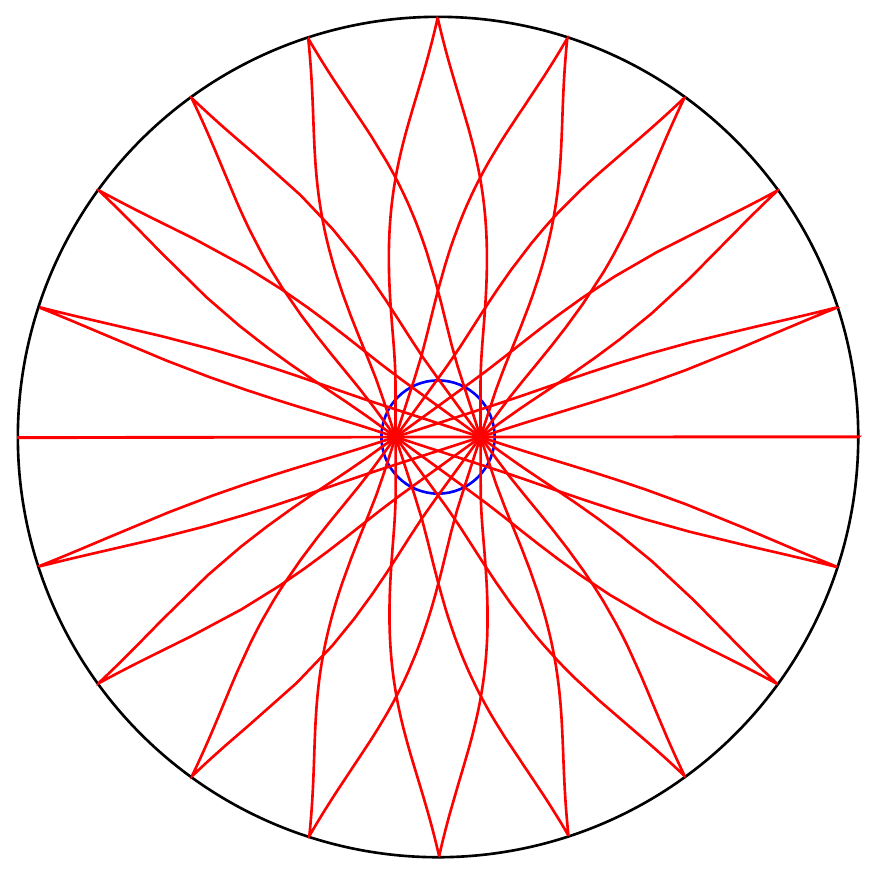}
\end{center}
\caption{Approximate perfect lens formed by surrounding the medium from
  Fig.~\ref{rays} with a spherical mirror of radius $R$ (shown in black).}
\label{rays-focusing}
\end{figure}

The key ingredient of our lens is a spherically symmetric refractive index
profile $n(r)$ that focuses all parallel rays within a sphere of, say, radius 1
and a constant refractive index to a point at a distance $R>1$ from the center
of the sphere (Fig.~\ref{rays}).  To see why this can be useful, imagine that a
spherical mirror is placed at the radius $R$ (Fig.~\ref{rays-focusing}). A ray
that emerges from a point P inside the unit sphere reaches the mirror, is
reflected and enters the sphere again. Because of the law of reflection and the
above focusing property, the ray after re-entering the sphere will be parallel
with the original ray and lie opposite to it from the center of the
sphere. Therefore it will pass approximately through the point P' that is
opposite of P, viewed from the center of the sphere. This shows that an
approximate perfect real image of point P is formed at point P'.

\begin{figure}
\begin{center}
  \includegraphics[width=8cm]{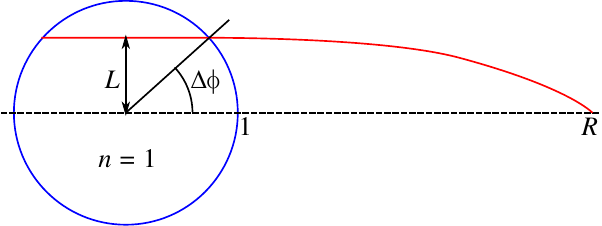}
\end{center}
\caption{Notation used for deriving the refractive index profile $n_R(r)$.}
\label{ray}
\end{figure}

To find the corresponding refractive index $n(r)$, we use the expression for
the polar angle swept by the ray during propagation from $r=1$ to $r=R$
\cite{Landau-shorter} (see Fig.~\ref{ray}):
\begin{equation}
  \Delta\phi=L\int_1^{R}\frac{\dd r}{r\sqrt{n^2r^2-L^2}}.
\label{deltaphi}
\end{equation}
Here $L=nr\sin\alpha$ and $\alpha$ denotes the angle between the ray and the
radius vector.  $L$ corresponds to the angular momentum in the equivalent
mechanical problem~\cite{Ulf-Thomas-book} and is conserved in any
spherically-symmetric refractive index profile. Assuming that the refractive
index inside the unit sphere is equal to one, $L$ is equal to the distance of
the ray from the center and $\Delta\phi=\arcsin L$ (Fig.~\ref{ray}).  Inserting
this into Eq.~(\ref{deltaphi}) and making the substitutions $r=\e^x$, $N=nr$, we
obtain an integral equation
\begin{equation}
L\int_0^{X} \frac{\mathrm{d}x}{\sqrt{N(x)^2-L^2}} = \arcsin{L},
\label{integral_equation0}
\end{equation}
where $X=\ln R$.  As has been shown~\cite{NJP}, this equation does not have an
exact solution; we therefore employed two different methods to find an
approximate solution $N(x)$ numerically.

In the first method, we changed the integration variable in
Eq.~(\ref{integral_equation0}) from $x$ to $N$ to obtain
\begin{equation}
L\int^{N_1}_1 u(N) \frac{\mathrm{d}N}{\sqrt{N^2-L^2}} = \arcsin{L},
\label{integral_equation}
\end{equation}
where $u(N)=\dd x/\dd N$ and $N_1=Rn(R)$. This equation can be solved by
Galerkin's method for the linear integral equations of the first kind
\cite{Polyanin} as follows. The unknown function $u(N)$ is first expanded as
$u(N) = \sum_i A_i \varphi_i (N)$, where $A_i$ are real coefficients and
$\varphi_i (N)$ is a set of functions on the interval $(1,N_1)$. Substituting
this expression into Eq.~(\ref{integral_equation}) and interchanging the
summation and integration, we obtain
\begin{equation}
\sum_i A_i g_i (L) = \arcsin{L}\,,
\label{first_sum}
\end{equation}
where $g_i (L) =L\int^{N_1}_1\varphi_i(N)(N^2-L^2)^{-1/2}\,\mathrm{d}N$.  For a
chosen set of functions $\varphi_i (N)$ we need to calculate the unknown
coefficients $A_i$. For this purpose we define another set of functions
$\psi_j(L)$ on the interval $(0,1)$. Multiplying both sides of
Eq.~(\ref{first_sum}) by $\psi_j(L)$, integrating over $L$ from $0$ to $1$ and
interchanging the order of summation and integration, we obtain the matrix
equation
\begin{equation}
\sum_i \sigma_{ij} A_i = B_j\,,
\label{matrix_equation}
\end{equation}
where $\sigma_{ij} = \int^1_0 \psi_j (L)g_i(L)\, \mathrm{d}L$ and $B_j =
\int^1_0 \psi_j (L) \arcsin{L}\, \mathrm{d}L$.  The unknown coefficients $A_i$
are then solutions of the system of linear equations
(\ref{matrix_equation}). Using the calculated coefficients $A_i$, the
approximate solution of function $u(N)$ is finally obtained.  In our
calculation, we have chosen polynomials as basis functions, $\varphi_i (N) =
N^i,\psi_j (L) = L^j$, with $i,j$ running from $0$ to some maximum value $M$.

The second, less sophisticated but equally efficient method, was based on
numerical minimization of the lens aberration. We have represented the function
$N(x)$ in Eq.~(\ref{integral_equation0}) by a polynomial $N(x)=1+\sum_1^k a_i
x^i$ with coefficients $a_i$ and calculated the aberration
\begin{equation}
A=\sum_{\{L_i\}}\left(L_i\int_0^{X}
\frac{\mathrm{d}x}{\sqrt{N(x)^2-L_i^2}}-\arcsin{L_i}\right)^2
\end{equation}
as a function of $a_1,\dots,a_k$. Here $L_i$ denotes a chosen set of
representative values of $L$ from the interval $(0,1]$. To find the minimum
  aberration, we have employed the numerical function NMinimize of the program
  Mathematica. It turned out that using polynomial of degree $k=5$ and ten
  uniformly distributed values of $L_i$ gave a refractive index with a
  negligible aberration; for example, for $X=3$ the mean difference of the
  angle $\Delta\phi$ from the correct value was $10^{-6}$ radians.
It turned out that the method works very well for $X\ge2$, but does not work
for $X\le1$, which would correspond to $R\le \rm e$. 
\begin{figure}
\begin{center}
  \includegraphics[width=8cm]{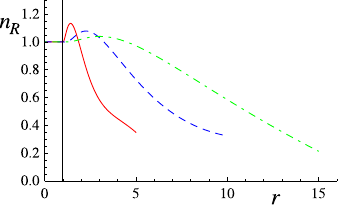}
\end{center}
\caption{Refractive index $n_R(r)$ calculated by the method of minimizing
  aberration for $R=5$ (red full line), $R=10$ (blue dashed line) and $R=15$
  (green dash-dotted line). The vertical line marks the border of the
  homogeneous region at $r=1$.}
\label{n(r)}
\end{figure}

Both methods give similar results for $n(r)$.  Fig.~\ref{n(r)} shows the
refractive index $n(r)$ obtained by the second method for several values of
$R$. The ray trajectories are shown in Fig.~\ref{rays} for $R=7$.  Of course,
the whole function $n(r)$ (including the region $r<1$) can be multiplied by any
real constant $C$ without changing the focusing properties of the lens. In
addition, the size of the lens can be scaled by an arbitrary factor $D$. If we
denote the original refractive index distribution by $n_{R}(r)$ instead of
$n(r)$ to emphasize that it depends on the parameter $R$, then the most general
refractive index of our lens becomes $n(r)=Cn_{R}(r/D)$. This lens focuses
parallel rays within the sphere of radius $D$ to the point at the distance $DR$
from its center. Using for instance $R=7$ and setting $C=2.74$ to bring $n(r)$
above one in the whole lens, we get $1\le n\le3.11$, which is a very moderate
range. 

\begin{figure}
\vspace{3mm}
\begin{center}
  \includegraphics[width=7cm]{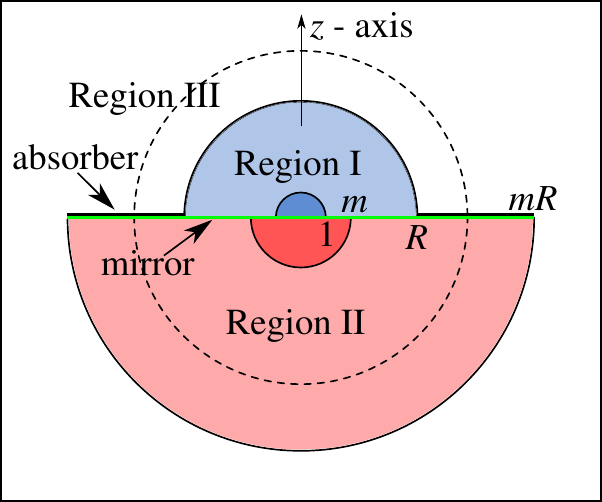}
\end{center}
\caption{Regions in the approximate magnifying perfect lens, the $z$ axis
  ($\theta=0$) being vertical. The radii of the individual spheres are marked
  along horizontal axis. The dashed line shows the radius $a=R\sqrt m$ of
  Maxwell's fish eye profile located in Region III. The object and image spaces
  are marked by a slightly darker color.}
\label{regions}
\end{figure}

Having described the lens that approximately focuses parallel rays to a single
point, we proceed now to construction of the magnifying lens. For this purpose,
we divide the whole Euclidean 3D space into three regions denoted I, II and III
(see Fig.~\ref{regions}).  Using spherical coordinates $(r,\theta,\fii)$
centered at a point O, we define the regions as follows. Region I is given by
the conditions $0\le r\le R$ and $0\le\theta<\pi/2$; region II is given by the
conditions $0\le r\le mR$ and $\pi/2<\theta\le\pi$, where $m\ge1$ is going to
be the lens magnification; region III occupies the rest of the space.
Refractive index in region I is given by the above described profile, i.e.,
$n_{\rm I}=n_{R}(r)$.  Refractive index in region II is $n_{\rm
  II}=n_{R}(r/m)/m$.  Refractive index in region III is described by
Maxwell's fish eye profile 
\begin{equation}
n_{\rm III}=\frac{n_1(1+1/m)}{1+r^2/(mR^2)},
\end{equation}
with the fish eye radius, i.e., the radius of ray that runs at a fixed distance
from the center O, equal to $a=R\sqrt m$, and $n_1=n_R(R)$. It is easy to check
that the refractive index is continuous at the hemispherical interfaces of
region III with both regions I and II, i.e., $n_{\rm I}(R)=n_{\rm III}(R)$ and
$n_{\rm II}(mR)=n_{\rm III}(mR)$.

The radius $a$ of Maxwell's fish eye was chosen such that it images the
hemispheric interface of regions I and III to the hemispheric interface of
regions II and III. Indeed, the relation of radial coordinates $r,r'$ of a
point and its image in the fish eye is $rr'=a^2$~\cite{BornWolf}, which is satisfied in
our case.

\begin{figure}
\begin{center}
\includegraphics[angle=0,width=7.5cm]{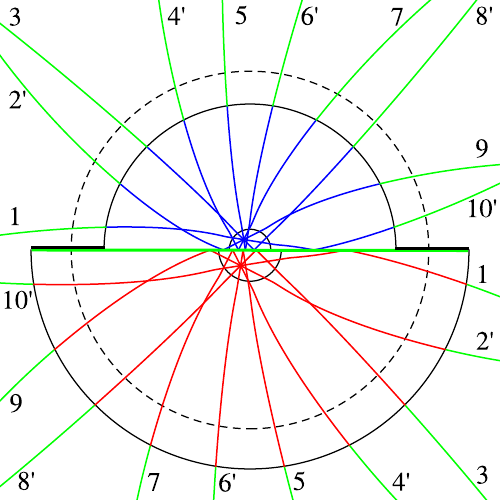}
\end{center}
\caption{Ray tracing in the approximate magnifying perfect lens with
  $m=3/2$. The parts of the rays in regions I, II and III are shown in blue,
  red and green, respectively. The numbers help to match the two parts of the
  same ray. The primes mark the rays that are reflected from the mirror before
  entering region III while unprimed rays are reflected after leaving region
  III.}
\label{full_lens}
\end{figure}

We will show now that this device approximately images the object space, which
is the homogeneous hemisphere of radius $r=1$ and refractive index $n=1$ in
region I, to the image space, which is the homogeneous hemisphere with radius
$r=m$ and refractive index $n=1/m$ in region II (see
Fig.~\ref{regions}). Consider a ray emerging from some point P placed at the
radius vector $\boldsymbol r_{\rm P}$ in the object space in the direction
described by a unit vector $\boldsymbol\nu$ that has a positive component in
$z$-direction. Due to the focusing properties of the profile $n_R(r)$ in region
I, it will hit the interface of regions I and III at the point A$\,=
R\boldsymbol\nu$.  The Maxwell's fish eye profile in region III will ensure
that the ray will then propagate along a circle and hit the interface of
regions III and II at the point B$\,=-mR\boldsymbol\nu$. Moreover, since the
ray is a segment of a circle, clearly the angle between the ray and the
straight line AOB at point A is the same as the angle between the ray and the
line AOB at point B.  The ray hence enters region II with the same impact angle
with which it left region I. Now since the index profiles in regions I and II
differ only by a spatial scaling by $m$ and a multiplicative factor, it is
clear that the shape of the ray in region II will be the same as its shape in
region I, up to scaling by $m$. In particular, when the ray enters the image
space, it will resume its original propagation direction $\boldsymbol\nu$, but
the straight line on which it lies will be $m\times$ more distant from the
origin O than the straight line of the first segment of the ray in region I.
This means that the ray will be directed towards the point P' placed at
$m\boldsymbol r_{\rm P}$.  Now if many rays emerge from point P, they will form
an image at P'; however, this point lies outside of region II, so the image
would be virtual. Because of this, we place a mirror at the interface of
regions I and II that changes the virtual image at P' into a real image at a
point P'' at position $\hat Z m\boldsymbol r_{\rm P}$, with $\hat Z$ meaning
the operation of inverting the $z$-coordinate.  Furthermore, if we make the
mirror double-sided, then we can take advantage also of the rays that emerge
``down'' from P (those with negative $z$-component of $\boldsymbol\nu$). These
rays will be reflected from the mirror while still propagating in object space
and then reach the image P'' without a further reflection, as marked in
Fig.~\ref{full_lens} by primes.  It turns out, however, that some rays with
positive $z$-component of $\boldsymbol\nu$ will hit the flat interface between
regions III and II.  Since it is not possible to use these rays for imaging, we
block them by placing an absorber at this flat interface on the side of region
III, keeping the mirror on the side of region II.  This way, not all rays
emerging from P reach P'', but it is a negligible minority of rays that are
lost in this way, especially if the magnification $m$ is not too large. Still,
it is not necessary for a device to capture all rays to be called a perfect
lens within geometrical optics~\cite{BornWolf}.

In summary, we have proposed a lens that makes an approximate magnified perfect
real image of a homogeneous 3D region. Performance of this lens in the full
wave optics regime is a subject of investigation, but we believe that it may
provide sub-wavelength resolution similarly as Maxwell's fish eye.
Manufacturing this lens would be difficult because the refractive index of
Maxwell's fish eye profile in region III goes to zero for $r\to\infty$, and
rays forming the image do get very far from the origin. Multiplying the whole
index profile by a large number will not help much because then the index in
the object and the image spaces then becomes very large. An option how to
reduce refractive index range significantly would be to position regions I and
II differently and use some other, not spherically symmetric index profile in
region III to image the surface of one sphere to the other. On the other hand,
the index profile $n_R(r)$ that is a part of the device can work as an
approximate perfect lens on its own, requiring just a moderate refractive index
range and therefore having potential to become a practical device.

After we published this paper on the arXiv, our colleague Klaus Bering proved
analytically that there exists no function $N(x)$ that solves the integral
equation~(\ref{integral_equation0}) exactly and therefore the refractive index
$n_R$ with the focusing properties does not exist either. Before that, we
believed that Eq.~(\ref{integral_equation0}) does have an exact solution but
that we just could not find it. A detailed proof of the non-existence of the
solution is given in Ref.~\cite{NJP}. In spite of that, we believe that even
though our device works only approximately, it still may have value for
further research; therefore we have replaced the paper on arXiv instead of
withdrawing it. In the meantime, we have found another method for designing
magnifying perfect lenses (or absolute instruments)~\cite{PRA} that works
exactly and is much more flexible than the method presented in this paper.

We thank Klaus Bering, Ulf Leonhardt, Michael Krbek and Aaron Danner for very
useful discussions.

%\bibliography{bibtex-tom}{}

\begin{thebibliography}{10}
\expandafter\ifx\csname bibnamefont\endcsname\relax
  \def\bibnamefont#1{#1}\fi
\expandafter\ifx\csname bibfnamefont\endcsname\relax
  \def\bibfnamefont#1{#1}\fi
\expandafter\ifx\csname url\endcsname\relax
  \def\url#1{\texttt{#1}}\fi
\expandafter\ifx\csname urlprefix\endcsname\relax\def\urlprefix{URL }\fi
\providecommand{\bibinfo}[2]{#2}
\providecommand{\eprint}[2][]{\url{#2}}

\bibitem{Pendry2000}
\bibinfo{author}{\bibfnamefont{J.~B.} \bibnamefont{Pendry}},
  \bibinfo{journal}{Phys. Rev. Lett.} \textbf{\bibinfo{volume}{85}},
  \bibinfo{pages}{3966} (\bibinfo{year}{2000}).

\bibitem{Shalaev2007-review}
\bibinfo{author}{\bibfnamefont{V.~M.} \bibnamefont{Shalaev}},
  \bibinfo{journal}{Nature Photonics} \textbf{\bibinfo{volume}{1}},
  \bibinfo{pages}{41} (\bibinfo{year}{2007}).

\bibitem{Fang2005-neg_n-silver_imaging}
\bibinfo{author}{\bibfnamefont{N.}~\bibnamefont{Fang}},
  \bibinfo{author}{\bibfnamefont{H.}~\bibnamefont{Lee}},
  \bibinfo{author}{\bibfnamefont{C.}~\bibnamefont{Sun}}, \bibnamefont{and}
  \bibinfo{author}{\bibfnamefont{X.}~\bibnamefont{Zhang}},
  \bibinfo{journal}{Science} \textbf{\bibinfo{volume}{308}},
  \bibinfo{pages}{534} (\bibinfo{year}{2005}).

\bibitem{Ulf2010-fisheye}
\bibinfo{author}{\bibfnamefont{U.}~\bibnamefont{Leonhardt}} \bibnamefont{and}
  \bibinfo{author}{\bibfnamefont{T.~G.} \bibnamefont{Philbin}},
  \bibinfo{journal}{Phys. Rev. A} \textbf{\bibinfo{volume}{81}},
  \bibinfo{pages}{011804(R)} (\bibinfo{year}{2010}).

\bibitem{Ulf2009-fisheye}
\bibinfo{author}{\bibfnamefont{U.}~\bibnamefont{Leonhardt}},
  \bibinfo{journal}{New J. Phys.} \textbf{\bibinfo{volume}{11}},
  \bibinfo{pages}{093040} (\bibinfo{year}{2009}).

\bibitem{Ma2010-fisheye}
\bibinfo{author}{\bibfnamefont{Y.~G.} \bibnamefont{Ma}},
  \bibinfo{author}{\bibfnamefont{C.}~\bibnamefont{Ong}},
  \bibinfo{author}{\bibfnamefont{S.}~\bibnamefont{Sahebdivan}},
  \bibinfo{author}{\bibfnamefont{T.}~\bibnamefont{Tyc}}, \bibnamefont{and}
  \bibinfo{author}{\bibfnamefont{U.}~\bibnamefont{Leonhardt}}
  (\bibinfo{year}{2010}), \eprint{arXiv:1007.2530}.

\bibitem{BornWolf}
\bibinfo{author}{\bibfnamefont{M.}~\bibnamefont{Born}} \bibnamefont{and}
  \bibinfo{author}{\bibfnamefont{E.}~\bibnamefont{Wolf}},
  \emph{\bibinfo{title}{Principles of optics}} (\bibinfo{publisher}{Cambridge
  University Press}, \bibinfo{year}{2006}).

\bibitem{Minano2006}
\bibinfo{author}{\bibfnamefont{J.~C.} \bibnamefont{{Mi\~nano}}},
  \bibinfo{journal}{Opt. Express} \textbf{\bibinfo{volume}{14}},
  \bibinfo{pages}{9627} (\bibinfo{year}{2006}).

\bibitem{Luneburg1964}
\bibinfo{author}{\bibfnamefont{R.~K.} \bibnamefont{Luneburg}},
  \emph{\bibinfo{title}{Mathematical Theory of Optics}}
  (\bibinfo{publisher}{University of California Press, Berkeley},
  \bibinfo{year}{1964}).

\bibitem{Landau-shorter}
\bibinfo{author}{\bibfnamefont{L.~D.} \bibnamefont{Landau}} \bibnamefont{and}
  \bibinfo{author}{\bibfnamefont{E.~M.} \bibnamefont{Lifshitz}},
  \emph{\bibinfo{title}{A shorter course of theoretical physics}}
  (\bibinfo{publisher}{Pergamon Press}, \bibinfo{year}{1972}).

\bibitem{Ulf-Thomas-book}
\bibinfo{author}{\bibfnamefont{U.}~\bibnamefont{Leonhardt}} \bibnamefont{and}
  \bibinfo{author}{\bibfnamefont{T.}~\bibnamefont{Philbin}},
  \emph{\bibinfo{title}{Geometry and Light: The Science of Invisibility}}
  (\bibinfo{publisher}{Dover, Mineola}, \bibinfo{year}{2010}).

\bibitem{NJP} T. Tyc, L. Herz\'anov\'a, Martin \v Sarbort and Klaus Bering, New
  J. Phys. {\bf 13}, 115004 (2011).

\bibitem{Polyanin}
\bibinfo{author}{\bibfnamefont{A.}~\bibnamefont{Polyanin}} \bibnamefont{and}
  \bibinfo{author}{\bibfnamefont{A.}~\bibnamefont{Manzhirov}},
  \emph{\bibinfo{title}{Handbook of integral equations}}
  (\bibinfo{publisher}{Chapman \& Hall}, \bibinfo{year}{2008}).

\bibitem{PRA} T. Tyc, Phys. Rev. A 84, 031801(R) (2011).

\end{thebibliography}
%\bibliographystyle{apsrev}

\end{document}